# Production of $\chi_c$ and $\eta_c$ production in ultra-peripheral collisions with two-photon processes

Gongming Yu[1], Yanbing Cai[2], Quangui Gao[3], and Qiang Hu[4]

[1]Fundamental Science on Nuclear Safety and Simulation Technology Laboratory, Harbin Engineering University, Harbin 150000, China
[2] Guizhou Key Laboratory in Physics and Related Areas, and Guizhou Key Laboratory of Big Data Statistic Analysis, Guizhou University of Finance and Economics, Guiyang 550025, China
[3]Department of Physics, Yuxi Normal University, Yuxi 653100, China
[4]Institute of Modern Physics, Chinese Academy of Sciences, Lanzhou 730000, China

We calculate the production of $\chi_c$ and $\eta_c$ by the semi-coherent and coherent two-photon interaction in ultra-peripheral heavy ion collisions at Relativistic Heavy Ion Collider (RHIC) and Large Hadron Collider (LHC) energies. The differential cross section of transverse momentum distribution and rapidity distribution for $AB \xrightarrow{\gamma\gamma} AHB$ ($H=\chi_c$, and $\eta_c$) are estimated by using the equivalent photon approximation in ultra-peripheral nucleus-nucleus collisions. The numerical results demonstrate that the experimental study of $\chi_c$ and $\eta_c$ in ultra-peripheral nucleus-nucleus collisions is feasible at RHIC and LHC energies.

Key words: Charmonium, two-photon processes, ultra-peripheral heavy ion collisions

PACS: 13.85.Ni, 14.40.Pq, 25.60.Bx, 25.75.-q

## I. INTRODUCTION

The dominant processes in ultra-peripheral heavy ion collisions are two-photon interaction in the equivalent photon approximation with large impact parameter. The equivalent photon method which treated electromagnetic fields of a moving charged particle as a flux of qusi-real photons proposed by Enrico Fermi [1]. Consequently, Weizsacker and Williams applied this method to relativistic nucleus [2,3]. The equivalent photon flux presented in ultraperipheral collisions, that the two ions interact via their cloud of qusi-real photons, become very high at Relativistic Heavy Ion Collider (RHIC) and Large Hadron Collider (LHC) energies, and has been found many useful applications in heavy quarkonium production [4-14]. In recent years, the measurements for heavy quarkonium photoproduction have been reported by PHENIX Collaboration [15] at RHIC, as well as ALICE Collaboration [16-19] at LHC.

Despite considerable efforts both in experiment and theory, the heavy quarkonium production mechanism in ultraperipheral collisions is still not fully understood. In the present work, we investigate the semi-coherent two-photon interaction processes for large-$p_T$ $\chi_c$ and $\eta_c$ production, as well as rapidity distributions with coherent two-photon interaction processes in nucleus-nucleus collisions at RHIC and LHC energies. In calculations of ultraperipheral nucleus-nucleus collisions, the impact parameter is usually required to be larger than the sum of the two nuclear radii, b > $R_A$ + $R_B$, and the radiated photons interact with each other. The photons of ions are coherently radiated by the whole nucleus, since the limit on the minimum photon wavelength is greater than the nuclear radius. In the transverse plane with no Lorentz contraction, an upper limit on the transverse momentum of the photon emitted by nucleus A is $p_T \leq \hbar c/R_A$, since the uncertainty principle. In the longitudinal direction, the maximum possible momentum is $p_T \leq \hbar \gamma_L c/R_A$, that is multiplied by a Lorentz factor ($\gamma_L$) due to the Lorentz contraction of the nucleus. Consequently, the maximum energy for γγ collision in a symmetric nucleus-nucleus collision is $2\hbar\gamma_L c/R_A$, that is large enough for studying the heavy quarkonium at RHIC and LHC.

In this paper, we report a feasibility study of the coherent and semi-coherent two-photon production process for $\chi_c$ and $\eta_c$ at RHIC and LHC. In Section II, we present coherent and semi-coherent two-photon processes for $\chi_c$ and $\eta_c$ at RHIC and LHC energies. The numerical results for nucleus-nucleus collisions at RHIC and LHC energies are plotted in Section III. Finally, the conclusion is given in Section IV.

## II. GENERAL FORMALISM

According to the equivalent photons approximation, the intensity of the electromagnetic field, and therefore the number of photons in the cloud surrounding the nucleus, is proportional to $Z^2$. Thus the two photon interactions are highly favored in heavy ions collisions. The differential cross-section for the $\chi_c$ and $\eta_c$ production from the semi-coherent two-photon process in ultra-peripheral nucleus-nucleus collisions can be written as

$$d\sigma\left(AB \xrightarrow{\gamma\gamma} AHB\right) = \hat{\sigma}_{\gamma\gamma \to H}(W) dN_1(\omega_1) dN_2(\omega_2), \tag{1}$$

where the energies for the photons emitted from the nucleus are $\omega_{1,2} = \frac{W}{2}\exp(\pm y)$, with $W^2 = 4\omega_1\omega_2$, and the transformations $d\omega_1 d\omega_2 = (W/2)dWdy$ can be performed.

The total cross section $\hat{\sigma}_{\gamma\gamma \to H}(W)$ for the $\chi_c$ and $\eta_c$ production can be written in terms of the two-photon decay width of the corresponding state as [20-24]

$$\hat{\sigma}_{\gamma\gamma \to H}(W) = 8\pi^2(2J+1)\frac{\Gamma_{H \to \gamma\gamma}}{M}\left(W^2 - M^2\right), \tag{2}$$

here $J$ and $M$ are the spin and mass of the produced $\chi_c$ and $\eta_c$, respectively. The two-photon decay width $\Gamma_{H \to \gamma\gamma}$ for the $\chi_c$ and $\eta_c$ can be taken from the experiment [25].

In the equivalent photons approximation, the flux of photons from the two relativistic nuclei with Z times the electric charge moving with a relativistic factor $\gamma \gg 1$, which is respect to some observer develops an equally strong magnetic-field component. Then the equivalent photon spectra for the relativistic nucleus can be obtained as [26-29]

$$\frac{dN(\omega, q^2)}{d\omega} = \frac{Z^2\alpha}{\pi\omega}\int d^2q_T \frac{q_T^2}{\left(q_T^2 + \omega^2/\gamma^2\right)^2} F_N^2(q_T^2 + \omega^2/\gamma^2), \tag{3}$$

where ω is the photon energy, γ is the relativistic factor, $F_N(q^2)$ is the nuclear form factor of the equivalent photon source, and $q^2 = (q_T^2 + \omega^2/\gamma^2)^2$ is the momentum transfer of the relativistic nuclei projectile.

For a realistic nucleus, the form factor [30] can be considered as a convolution of the hard sphere with radius $R_T$ and Yukawa potential,

$$F_N(q) = \frac{4\pi d_0}{Aq^3}\left[\sin(qR_A) - qR_A\cos(qR_A)\right]\left(\frac{1}{1+a^2q^2}\right), \tag{4}$$

where the parameters $d_0=0.13815$ fm$^{-3}$, $R_T=1.2A^{1/3}$ fm, and $a=0.7$ fm can be found in Ref. [31], and $A$ is the nucleon number.

In the semi-coherent two-photon process, the total transverse momentum of $\chi_c$ and $\eta_c$ meson are $\boldsymbol{p}_T = \boldsymbol{q}_{1T} + \boldsymbol{q}_{2T} \approx \boldsymbol{q}_{1T}$, since the momentum for photons are $q_1=(\omega_1, \boldsymbol{q}_{1T}, q_{1z})$ and $q_2=(\omega_2, \boldsymbol{q}_{2T}, q_{2z})$, where $\boldsymbol{q}_{iT}$ is the transverse momentum of the $i$-th photon. Consequently, the differential cross section for the ultra-peripheral collisions can be written as

$$\frac{d\sigma}{d^2 p_T dy} = \frac{8Z^4\alpha^2}{\pi^2}(2J+1)\frac{\Gamma_{H\to\gamma\gamma}}{M^3}\frac{F_N^2\left(p_T^2+\omega_1^2/\gamma^2\right)}{p_T^2} \\ \times \int d^2 q_T q_{2T}^2 \frac{F_N^2\left(q_{2T}^2+\omega_2^2/\gamma^2\right)}{\left(q_{2T}^2+\omega_2^2/\gamma^2\right)^2}, \quad (5)$$

where $\gamma$ is the relativistic factor, and the transverse momentum of photon is $q_{2T} > 0.2$ GeV due to the single track acceptance condition [26].

If we consider the effects of strong absorption, which implies that hadronic interactions will dominate in relativistic electromagnetic interactions, the total number of photons from a ultra-peripheral collisions can be obtained by integrating over all impact parameters larger than the nuclear radius. Consequently, the cross sections with considering the accurate hadronic interaction probabilities for the equivalent two-photon luminosity in the impact parameter space can be written as

$$d\sigma\left(AB \xrightarrow{\gamma\gamma} AHB\right) = \hat{\sigma}_{\gamma\gamma\to H}(W)dN_1(\omega_1,\vec{b}_1)dN_2(\omega_2,\vec{b}_2)S_{abs}^2(\vec{b}), \quad (6)$$

where the photon spectrum with the charge form factor $F_N(q^2)$ as follows [32-36]

$$\frac{dN(\omega,\vec{b})}{d\omega d^2 b} = \frac{Z^2\alpha}{\pi^2\omega}\left[\int_0^\infty dq_T q_T^2 \frac{F_N^2\left(q_T^2+\omega^2/\gamma^2\right)}{q_T^2+\omega^2/\gamma^2}J_1(bq_T)\right]^2, \quad (7)$$

here $J_1(x)$ is Bessel function.

The absorptive factor $S_{abs}^2(\vec{b})$ can be expressed in terms of the probability of interaction between the nuclei with a given impact parameter [37,38],

$$S_{abs}^2(\vec{b}) = 1 - P_H(\vec{b}), \quad (8)$$

with

$$P_H(\vec{b}) = 1 - \exp\left[\sigma_{NN} T_{AA}\right] \\ = 1 - \exp\left[\sigma_{NN}\int d^2 r T_A(\vec{r})T_A(\vec{r}-\vec{b})\right], \quad (9)$$

where $T_A$ is nuclear thickness function [39], and $\sigma_{NN}$ being the total hadronic interaction cross section, 52mb at RHIC and 88mb at LHC [40].

Based on the transformations $d\omega_1 d\omega_2=(W/2)dWdy$, the rapidity distributions for the coherent two-photon process can be written in the terms of the rapidity for $\chi_c$ and $\eta_c$ meson as the following

$$\frac{d\sigma}{dy} = 8\pi^2 (2J+1) \frac{\Gamma_{(l^+l^-)\to\gamma\gamma}}{M^3} \int d^2b_1 d^2b_2 S_{abx}^2(\vec{b})$$

$$\times \frac{Z^2\alpha}{\pi^2} \left[ \int_0^\infty dq_{1T} q_{1T}^2 \frac{F_N^2\left(q_{1T}^2 + \omega_1^2/\gamma^2\right)}{q_{1T}^2 + \omega_1^2/\gamma^2} J_1(b_1 q_{1T}) \right]^2 \quad (10)$$

$$\times \frac{Z^2\alpha}{\pi^2} \left[ \int_0^\infty dq_{2T} q_{2T}^2 \frac{F_N^2\left(q_{2T}^2 + \omega_2^2/\gamma^2\right)}{q_{2T}^2 + \omega_2^2/\gamma^2} J_1(b_2 q_{2T}) \right]^2,$$

where $q_{iT}$ is the transverse momentum of the $i$-th photon.

### III. NUMERICAL RESULTS

The equivalent photon fluxes for the heavy nucleus become very large at the RHIC and LHC energies, since the photon flux scales as $Z^2$. This implies that the two-photon differential cross-section scales as $Z^4$. In Fig. 1, we plot the differential cross section for large-$p_T$ $\chi_{c0}$, $\chi_{c2}$, $\eta_c(1S)$, and $\eta_c(2S)$ production in the semi-coherent two-photon processes, that the transverse momentum of one of the photons become small, then the whole nucleus acts coherently without considering the effects of strong absorption. The rapidity distributions of $\chi_{c0}$, $\chi_{c2}$, $\eta_c(1S)$, and $\eta_c(2S)$ produced by the coherent two-photon approach in the impact parameter space are plotted in Fig. 2. The main sources of changes in the differential cross sections are the magnitude of the decay width and the spin of the produced particle ($\chi_{c0}$, $\chi_{c2}$, $\eta_c(1S)$, and $\eta_c(2S)$), since the difference of mass is small.

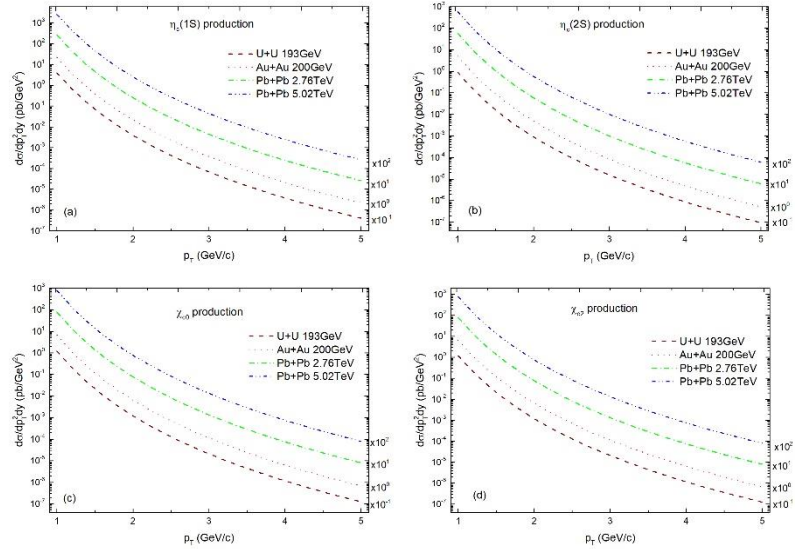

FIG. 1. The differential cross section for $\chi_c$ and $\eta_c$ from the semicoherent two-photon interaction (without impact parameter b) in ultraperipheral heavy ion collisions at RHIC and LHC. The dashed line (wine line) is for U+U collisions with $\sqrt{s_{NN}} = 193 GeV$, the dotted line (red line) is for Au+Au collisions with $\sqrt{s_{NN}} = 200 GeV$, the dashed-dotted line (green line) for Pb+Pb collisions with $\sqrt{s_{NN}} = 2.76 TeV$, the dashed-dotted-dotted line (blue line) is for Pb+Pb collisions with $\sqrt{s_{NN}} = 5.02 GeV$.

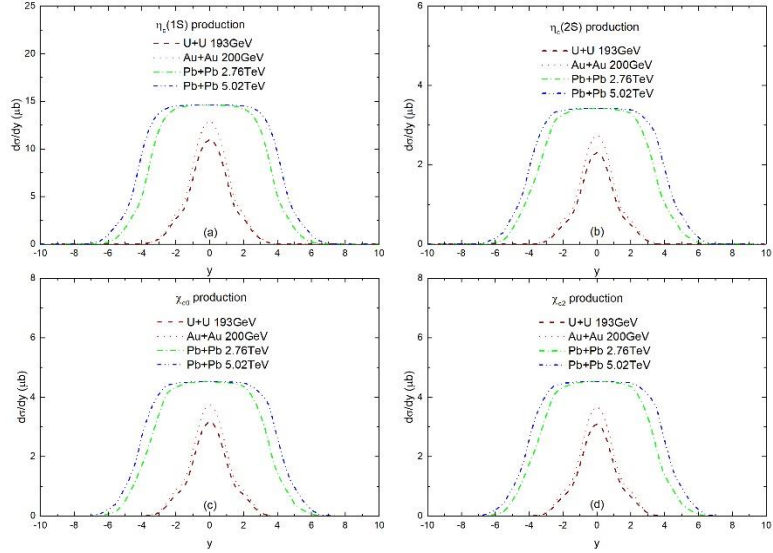

FIG. 2. The diffferential cross section for $\chi_c$ and $\eta_c$ production from the coherent two-photon interaction (with impact parameter $b$) in ultraperipheral heavy ion collisions at RHIC and LHC. The dashed line (wine line) is for U+U collisions with $\sqrt{s_{NN}} = 193 GeV$, the dotted line (red line) is for Au+Au collisions with $\sqrt{s_{NN}} = 200 GeV$, the dashed-dotted line (green line) for Pb+Pb collisions with $\sqrt{s_{NN}} = 2.76 GeV$, the dashed-dotted-dotted line (blue line) is for Pb+Pb collisions with $\sqrt{s_{NN}} = 5.02 GeV$.

IV. CONCLUSION

In summary, we have investigated the production of $\chi_c$ and $\eta_c$ from the semi-coherent and coherent two-photon processes with the equivalent photon approximation in ultra-peripheral collisions at RHIC and LHC energies. The charge distribution form factor and effects of strong absorption are considered in the two-photon interactions processes. Our calculations show that the differential cross sections for $\chi_c$ and $\eta_c$ produced in nucleus-nucleus collisions can not be neglected at the RHIC and LHC energies.

V. ACKNOWLEDGEMENTS

This work is supported by Heilongjiang Science Foundation Project under Grant No. LH2021A009, and National Natural Science Foundation of China under Grant No. 12063006.